\documentclass[12pt,a4paper]{article}
\usepackage{jheppub}

\title{Electroweak symmetry breaking by gravity}

\author[a,b]{Yuri Shtanov}

\affiliation[a]{Bogolyubov Institute for Theoretical Physics, \\ Metrologichna St.\@ 14-b, Kiev 03143, Ukraine} %
\affiliation[b]{Astronomical Observatory, Taras Shevchenko National University of Kyiv, \\ Observatorna St.\@ 3, Kiev 04053, Ukraine} %

\emailAdd{shtanov@bitp.kiev.ua}

\abstract{We consider a simple scale-invariant action coupling the Higgs field to the metric scalar curvature $R$ and containing an $R^2$ term that exhibits spontaneous breaking of scale invariance and electroweak symmetry. The coefficient of the $R^2$ term in this case determines the self-coupling of the Higgs boson in the Einstein frame, and the scalaron becomes a dilaton weakly coupled to the Higgs boson. Majorana mass terms for right-handed neutrinos can be generated in a scale-invariant manner by using the Higgs-field invariant; in this case, the existing experimental limits on the Higgs-boson total width rule out Majorana mass values in a certain range.  The model inherits the naturalness issues of general relativity connected with the smallness of the gravitational and cosmological constants.}


\arxivnumber{2305.17582}

\begin{document}

\maketitle
\flushbottom

\section{Introduction}

It has long been suggested in the literature that the global scale-invariance can be an exact symmetry of nature \cite{Fujii:1974bq, Fujii:1982ms, Englert:1975wj, Englert:1976ep, Wetterich:1987fk, Wetterich:1987fm, Shaposhnikov:2008xb, Shaposhnikov:2008xi, Gretsch:2013ooa, Shaposhnikov:2022dou, Shaposhnikov:2022zhj}.\footnote{In \cite{Englert:1975wj, Englert:1976ep}, this principle is extended to a local scale-invariance, or Weyl invariance.} Here, by global scale-invariance, we mean invariance of the Lagrangian density with respect to multiplication of the space-time metric and matter fields by appropriate constant factors (such transformations are often called dilatations). Lagrangians respecting this principle cannot contain dimensionful parameters (such as masses or gravitational and cosmological constants), hence, scale symmetry has to be dynamically broken in order to generate them. Since only the dimensionless ratios of such parameters are measured in experiment, it appears possible, in principle, to generate all dimensionful scales by a unique mechanism. Scale invariance can be maintained  on the quantum level, albeit at the expense of renormalisability \cite{Englert:1976ep, Shaposhnikov:2008xb, Shaposhnikov:2008xi, Gretsch:2013ooa, Shaposhnikov:2022dou, Shaposhnikov:2022zhj}.

Technically, this idea is usually implemented by introducing a special scalar field $\chi$, sometimes referred to as the ``metron'' (its logarithm or, often, the field $\chi$ itself is also called the dilaton; see \cite{Wetterich:2019qzx} for a review), whose expectation value gives birth to all dimensionful parameters.  In particular, the gravitational constant is generated by the scale-invariant coupling of the field $\chi$ to the metric curvature scalar $R$ via $\xi \chi^2 R$ term, following the old idea due to Brans and Dicke \cite{Brans:1961sx, Dicke:1961gz}. 

A natural question arises whether the Higgs field of the Standard Model can play the role of such a scalar field breaking the scale-invariance. It turns out that, even in the presence of the scale and electroweak symmetry-breaking potential, such a theory is not realistic: the coupling constant $\xi$ is required to be extremely large (see below), in which case the Higgs field becomes effectively massless and decouples from the other fields of the Standard Model \cite{Zee:1978wi, vanderBij:1993hx, vanderBij:1994bv}.  

In this letter, we point out that such a theory is remedied by adding the $R^2$ term to the gravitational action, thus introducing a new scalar degree of freedom (eventually to become a dilaton). In this case, one can start with a scale-invariant action even without any potential for the Higgs field; both the scale invariance and the electroweak symmetry become naturally dynamically broken. This is most easily seen in the Einstein frame of the theory, in which the usual Standard Model action arises together with the Einstein gravity and a new scalar degree of freedom (the dilaton) coupled to the Higgs field.  The scale-invariance of the theory in the original frame becomes the dilaton shift symmetry in the Einstein frame.  The dimensionless constant in front of the $R^2$ term regulates the self-coupling of the Higgs field, hence, the mass of the Higgs boson in this frame. On the other hand, the scale-invariant term $\lambda_\Phi \left( \Phi^\dagger \Phi \right)^2$ in the original action is responsible for the cosmological constant, hence, requires a tiny constant $\lambda_\Phi$. The extreme largeness of $\xi$ and smallness of $\lambda_\Phi$ represent fine-tuning issues in the original frame, equivalent to the naturalness issues of the Planck and  cosmological constants in the Einstein frame.

\section{The model}

Let us consider gravity coupled to the Higgs field of the Standard Model, with a scale-invariant total low-energy effective Lagrangian\footnote{We use the metric signature $(-, +, +, +)$ and system of units $\hbar = c = 1$.}
\begin{equation} \label{L}
L = \xi_* \Phi^\dagger \Phi R + \frac{\xi_*^2}{4 \lambda_*} R^2 - \lambda_\Phi \left( \Phi^\dagger \Phi \right)^2 - \left( D_\mu \Phi \right)^\dagger D^\mu \Phi + L_\text{m} \, ,
\end{equation} 
where $\xi_*$, $\lambda_*$ and $\lambda_\Phi$ are positive dimensionless constants, $\Phi$ is the Higgs doublet, and $D_\mu$ is the gauge covariant derivative involving the SU(2) and U(1) electroweak gauge fields and acting on the Higgs doublet $\Phi$. Note that the potential for the Higgs field does not contain the symmetry-breaking parameter, which is forbidden by the invariance of the Lagrangian density with respect to the global scaling $g_{\mu\nu} \to \zeta^2 g_{\mu\nu}$, $\Phi \to \zeta^{-1} \Phi$, etc., with $\zeta$ being a space-time constant. The part $L_\text{m}$ contains all the rest of matter fields, together with their couplings to the Higgs and gauge fields. It is assumed to be Weyl invariant, e.g., invariant with respect to local conformal transformations of the metric accompanied by appropriate local transformations of the matter fields; in particular, all couplings in $L_\text{m}$ are dimensionless. (This is the case in the Standard Model without the Majorana mass terms for right-handed neutrinos.  We will return to this issue below.)

Model of the form \eqref{L} with a special scalar field in place of the Higgs field was under investigation in \cite{Gorbunov:2013dqa, Rinaldi:2015uvu, Tambalo:2016eqr, Ghoshal:2022qxk} as a viable model of inflation. Inflation based on the Higgs field in \eqref{L} would not be realistic, as will be discussed in the end of Sec.~\ref{sec:discuss}.

Introducing an auxiliary scalar field $\chi_*$ with canonical dimension of mass, one can write Lagrangian \eqref{L} in the equivalent form
\begin{align} \label{L0}
L &= \xi_*^{} \chi_*^2 R - \lambda_* \left( \Phi^\dagger \Phi - \chi_*^2 \right)^2 - \lambda_\Phi \left( \Phi^\dagger \Phi \right)^2 - \left( D_\mu \Phi \right)^\dagger D^\mu \Phi + L_\text{m}  \\[3pt] &= \xi \chi^2 R - \lambda \left( \Phi^\dagger \Phi - \chi^2 \right)^2 - \lambda_\chi \chi^4 - \left( D_\mu \Phi \right)^\dagger D^\mu \Phi + L_\text{m} \, . \label{L1}
\end{align}
Indeed, finding the extremum of \eqref{L0} with respect to $\chi_*^2$ and substituting it into \eqref{L0} gives back equation \eqref{L}.  In equation \eqref{L1}, we have made a rescaling of the constants and of the auxiliary field according to
\begin{equation} \label{const}
\lambda = \lambda_* + \lambda_\Phi \, , \quad \lambda_\chi = \lambda_\Phi \left( 1 + \frac{\lambda_\Phi}{\lambda_*} \right) \, , \quad \xi = \xi_* \left( 1 + \frac{\lambda_\Phi}{\lambda_*} \right) \, , \quad \chi = \left( 1 + \frac{\lambda_\Phi}{\lambda_*} \right)^{-1/2} \chi_* \, .
\end{equation}

Theory \eqref{L1} is just the usual dilatonic extension of the Standard Model \cite{Fujii:1974bq, Fujii:1982ms} but without the kinetic term for the field $\chi$.  Note that, when $\lambda_\Phi = 0$, the Higgs-field potential is absent, we have also $\lambda_\chi = 0$, and the starred constants and auxiliary scalar field coincide with the unstarred ones. Expression \eqref{L1} continues to describe a scale-invariant theory, with the usual scaling of the auxiliary field $\chi \to \zeta^{-1} \chi$.

In the gravity sector, action \eqref{L1} is just the Brans--Dicke theory \cite{Brans:1961sx, Dicke:1961gz} with the field $\chi^2$ playing the role of the Brans--Dicke scalar that determines the gravitational coupling. As such, it has zero Brans--Dicke parameter $\omega$ (no kinetic term for the $\chi$ field), but has a non-trivial joint potential with the Higgs field. In the phase of unbroken scaling symmetry, we have $\chi = 0$ and $\Phi = 0$; this phase describes gravity at a special (singular) point, with infinite gravitational coupling, and does not correspond to the observed world. Just as in the Brans--Dicke theory, we need to assume that $\chi^2 \ne 0$ in the observable universe, so that the scaling symmetry is broken by the solution of the theory that describes our world\,---\,an idea realised in all such theories of broken scaling symmetry \cite{Fujii:1974bq, Fujii:1982ms, Englert:1975wj, Englert:1976ep, Wetterich:1987fk, Wetterich:1987fm, Shaposhnikov:2008xb, Shaposhnikov:2008xi, Shaposhnikov:2022zhj, Wetterich:2019qzx, Gorbunov:2013dqa, Rinaldi:2015uvu, Tambalo:2016eqr, Ghoshal:2022qxk, Ghilencea:2018dqd, Ghilencea:2021lpa, Shtanov:1994sh}. Scaling symmetry breaking in such theories is understood as a physical choice between the singular phase with unbroken symmetry and the phase with broken symmetry. The specific local value of $\chi^2 \ne 0$ does not matter because all such values are equivalent by scaling transformation. What really matters for the observable quantities is the relation between various dimensionless constants in the Lagrangian (see below).

Let us then consider the phase with broken scaling symmetry of theory \eqref{L1}, with the field $\chi$ acquiring a nonvanishing space-time value. For a constant (vacuum) expectation value of $\chi^2$, the vacuum expectation value of the Higgs-field invariant $\Phi^\dagger \Phi$ (which is the value minimising the Higgs-field potential in \eqref{L1}) becomes equal to $\chi^2$, while the factor $\xi \chi^2$ of the scalar curvature generates the squared Planck mass. The relation between the vacuum expectation value of $\chi^2$ and space-time curvature in this case is found from the equation of motion for $\chi^2$ by using the vacuum condition $\Phi^\dagger \Phi = \chi^2$:
\begin{equation} \label{pr}
\chi^2 \equiv \Phi^\dagger \Phi = \frac{\xi}{2 \lambda_\chi} R = \frac{\xi_*}{2 \lambda_\Phi} R \, ,
\end{equation}
where we have used the equality $\xi / 2 \lambda_\chi = \xi_* / 2 \lambda_\Phi$ that follows from \eqref{const}. Equation \eqref{pr} can also be obtained from \eqref{L} by finding the vacuum value of the Higgs field. A non-vanishing vacuum value of $\chi^2 \equiv \Phi^\dagger \Phi$ is then conditioned by a positive vacuum value of the scalar curvature $R$, implying the presence of an effective cosmological constant. In this sense, electroweak symmetry can be said to be broken by gravity.  On the other hand, in the limit $\lambda_\Phi \to 0$ (hence, also $\lambda_\chi \to 0$), we also have $R \to 0$ in the vacuum, while the value of $\chi^2 = \Phi^\dagger \Phi$ remains to be finite, setting the mass scale in the theory.

At this point, we note that related ideas of breaking scale and electroweak symmetry were realised in \cite{Ghilencea:2018dqd, Ghilencea:2021lpa} (see also \cite{Shtanov:1994sh}) for a Weyl-invariant theory with an additional gauge vector field; in this case, there arises a kinetic term for the field $\chi$ preserving the exact Weyl invariance of the theory.  In our case, Weyl invariance of the whole theory is not assumed; theory \eqref{L} is only globally scale-invariant and is much simpler by construction.

To elucidate the field dynamics in the phase with broken scaling symmetry, we proceed to the Einstein frame in a usual way. We parametrise the field $\chi^2 > 0$ by a new scalar field $\phi$ and make a Weyl rescaling of the metric:
\begin{equation} \label{om}
\chi^2 (\phi) = \frac{M^2}{2 \xi} \Omega^2 (\phi) \, , \qquad \Omega (\phi) = e^{\phi / \sqrt{6} M} \, , \qquad g_{\mu\nu} \to \Omega^{-2} g_{\mu\nu}\, .
\end{equation}
Here, $M$ is an arbitrary constant of dimension mass. We also exploit the Weyl invariance of the original action $S_\text{m}$ with Lagrangian $L_\text{m}$. This allows us to accompany the Weyl rescaling \eqref{om} by the corresponding Weyl rescaling of all other fields, including the Higgs field, which is transformed as
\begin{equation} \label{Htr}
\Phi \to \Omega \Phi \, .
\end{equation}  
Being Weyl invariant, the action $S_\text{m}$ retains its original form in terms of new fields. As a result of these transformations, Lagrangian \eqref{L1} in the Einstein frame becomes
\begin{align} \label{L2}
L = \frac{M^2}{2} \left( R - 2 \Lambda \right) - \frac12 \left( \partial \phi \right)^2 - \frac{1}{\sqrt{6} M} \partial_\mu \left( \Phi^\dagger \Phi \right) \partial^\mu \phi - \frac{1}{6 M^2} \Phi^\dagger \Phi\, \left( \partial \phi \right)^2 \nonumber \\[3pt] {}- \left( D_\mu \Phi \right)^\dagger D^\mu \Phi - \lambda \left( \Phi^\dagger \Phi - v^2 / 2 \right)^2 + L_\text{m} \, , 
\end{align}
where
\begin{equation} \label{Lambda}
v^2 = \frac{M^2}{\xi} \, , \qquad \Lambda = \frac{\lambda_\chi M^2}{4 \xi^2}  = \frac{\lambda_\chi v^4}{4 M^2} \, .
\end{equation}

We observe that the field $\phi$ has dropped out from the field potential, having become a dilaton, while the Higgs field acquired a standard electroweak symmetry-breaking potential. The second line in \eqref{L2} is just the Lagrangian of the Standard Model, while the first line is the Lagrangian for the Einstein gravity with the Planck mass $M$ and cosmological constant $\Lambda$, and for the dilaton $\phi$ with its derivative couplings to the Higgs field. These couplings are suppressed by inverse powers of the Planck mass $M$. The scale-invariance of the original action has transformed to the invariance of \eqref{L2} with respect to the shifts $\phi \to \phi + \text{const}$. 

The coupled Higgs--dilaton system in \eqref{L2} can be diagonalised by the transformation found for models of this type in \cite{Gorbunov:2013dqa, Tambalo:2016eqr, Ghoshal:2022qxk}. We choose the canonical unitary gauge for the electroweak SU(2) group, in which the Higgs doublet $\Phi$ takes the form
\begin{equation}\label{higgs}
\Phi = \frac{1}{\sqrt{2}} \begin{pmatrix} 0 \\ H \end{pmatrix} \, .
\end{equation}
After the transformation from $H$, $\phi$ to new fields ${\cal H}$, $\varphi$ by 
\begin{align}
H &= \sqrt{6} M \sinh \frac{{\cal H}}{\sqrt{6} M} \, , &{\cal H} &= \sqrt{6} M \text{arsinh\,} \frac{H}{\sqrt{6} M} \, , \label{H} \\[2pt]
\phi &= \varphi - \sqrt{6} M \ln \left( \cosh \frac{\cal H}{\sqrt{6} M} \right) \, , &\varphi &= \phi + \sqrt{\frac32} M \ln \left( 1 + \frac{H^2}{6 M^2} \right) \, ,
\end{align}
the Higgs--dilaton part in \eqref{L2} takes the form
\begin{equation}\label{Hd}
L_\text{H} = - \frac12 \left( \partial \varphi \right)^2 \cosh^2 \frac{\cal H}{\sqrt{6} M} - \frac12 \left( \partial {\cal H} \right)^2 - \frac{\lambda}{4} \left( 6 M^2 \sinh^2 \frac{\cal H}{\sqrt{6} M} - v^2 \right)^2 \, .
\end{equation}
This Lagrangian describes a massless dilaton $\varphi$ with kinetic term modified by the Higgs field, and a modified self-interacting Higgs field ${\cal H}$ with mass
\begin{equation}\label{hm}
m_h = \sqrt{2 \lambda} v \left( 1 + \frac{v^2}{6 M^2} \right)^{1/2} \, .
\end{equation}

Since the massless dilaton has only \emph{derivative\/} couplings, being a Goldstone boson related to the dynamical breaking of scale invariance, it evades the gravitational constraints \cite{Wetterich:1987fm, Shaposhnikov:2008xb} for the Brans--Dicke field \cite{Brans:1961sx}. In our case, this is also clear from Lagrangian \eqref{Hd}; it does not allow for a one-dilaton particle exchange, ensuring the absence of a long-range fifth force.

The values of all dimensionless constants in \eqref{L} are fixed by experiment. Indeed, the values of $M = \sqrt{1 / 8 \pi G} \approx 2.4 \times 10^{18}$~GeV and $v \approx 246$~GeV in the Standard Model imply
\begin{equation} \label{xi}
\xi = \frac{M^2}{v^2} \approx 10^{32} \, .
\end{equation}
The constant $\lambda$ determines the Higgs-boson mass \eqref{hm} $m_h \approx \sqrt{2 \lambda} v \approx 125$~GeV and self-coupling, and should be set to its established value $\lambda \approx 0.13$. It can be observed that the metric scalar curvature in Lagrangian \eqref{L} or \eqref{L1} enters in the combination $\xi R$ with a huge constant $\xi$ given by \eqref{xi}, whose large value is responsible for the weakness of gravity.

In view of \eqref{Lambda}, in order to account for the small observable value of $\Lambda \approx 4 \times 10^{-66}~\text{eV}^2$, the dimensionless constant $\lambda_\chi$ should be extremely small, $\lambda_\chi \approx 3 \times 10^{-56}$. Hence, $\xi_* \approx \xi$,  $\lambda_* \approx \lambda$ and $\lambda_\Phi \approx \lambda_\chi$ to a very high precision, which means that the self-coupling $\lambda$ of the Higgs field in \eqref{L2} or \eqref{Hd} is determined by the coefficient of the $R^2$ term in \eqref{L}. Given that the Higgs field has usual interactions with the matter fields in the original frame \eqref{L}, such an extreme smallness of $\lambda_\Phi$ represents a naturalness issue in this frame.  In the Einstein frame, it is translated to the naturalness issue for the cosmological constant. 

The Dirac masses of fermions in the Standard Model are generated by the usual Yukawa coupling terms, making the Dirac action Weyl invariant. Therefore, in all relevant terms in the Lagrangian, the Higgs field $H$ enters canonically, and just has to be expressed through ${\cal H}$ by \eqref{H}. To generate Majorana masses for right-handed neutrinos in a Weyl-invariant way in the original frame \eqref{L}, we can again exploit the Higgs field. 

The action for a Weyl spinor $\psi^A$, in the Penrose spinor-index notation \cite{Penrose:1987}, reads
\begin{equation}\label{ws}
S_\psi = \sqrt{2}\, {\rm i} \int \bar \psi^{A'} \nabla_{AA'} \psi^A \sqrt{- g}\, d^4 x \, ,
\end{equation}
where $\nabla_{AA'}$ is the spinor covariant derivative compatible with the space-time metric. Under the Weyl rescaling \eqref{om} of the metric, the spinor covariant derivative transforms as \cite{Penrose:1987}
\begin{equation}
\nabla_{AA'} \psi^B \to \nabla_{AA'} \psi^B - \epsilon_A{}^B \psi^C\, \nabla_{CA'} \ln \Omega \, .
\end{equation}
Here, $\epsilon_{AB}$ is the antisymmetric $\epsilon$-spinor field [it is transformed as $\epsilon_{AB} \to \Omega^{-1} \epsilon_{AB}$ under \eqref{om}]. Action \eqref{ws} is then Weyl invariant under simultaneous rescaling \eqref{om} of the metric and rescaling $\psi^A \to \Omega^2 \psi^A$ of the spinor. 

To construct a Weyl-invariant action for a Majorana spinor, we need a gauge-invariant scalar field that transforms canonically under \eqref{om}. With only the Higgs field of the Standard Model at our disposal, it is natural to use $\left( \Phi^\dagger \Phi \right)^{1/2}$ as such a scalar, and write the Lagrangian for the Majorana spinor in the form
\begin{equation}\label{sterile}
L_\psi = \sqrt{2}\, {\rm i}\, \bar \psi^{A'} \nabla_{AA'} \psi^A -  \frac{\gamma}{\sqrt{2}} \left( \Phi^\dagger \Phi \right)^{1/2} \left( \psi^A \psi_A + \bar \psi_{A'} \bar \psi^{A'} \right) \, ,
\end{equation}
where $\gamma$ is the coupling constant. The corresponding action is Weyl invariant and, after the electroweak symmetry breaking, the Majorana spinor acquires the mass $m_\psi = \gamma v$.  

Such a mechanism of mass generation implies interaction between the Higgs boson and Majorana fermion:
\begin{equation} \label{int}
L_\text{int} = - \frac{\gamma}{2} h \left( \psi^A \psi_A + \bar \psi_{A'} \bar \psi^{A'} \right) \, ,
\end{equation}
where the Higgs field $h$ is the deviation of ${\cal H}$ from its vacuum value in \eqref{Hd}, and interaction \eqref{int} is written to the leading order in $h / M$. Fermions with mass smaller than $m_h/2$ contribute to the width of the Higgs boson: 
\begin{equation}
\Gamma_{h \to \psi \psi} = \frac{\gamma^2}{16 \pi m_h^2} \left( m_h^2 - 4 m_\psi^2 \right)^{3/2} = \frac{m_\psi^2}{16 \pi v^2 m_h^2} \left( m_h^2 - 4 m_\psi^2 \right)^{3/2} \, . 
\end{equation}
The theoretical value of the total width of the $125$-GeV Higgs boson in the Standard Model is $\Gamma_h = 4.1$~MeV \cite{LHCHiggs:2016ypw}.  In order that the total width in our model be within the experimental limits $\Gamma_h = 4.5^{+ 3.3}_{- 2.5}$~MeV \cite{ATLAS:2023dnm}, for one such fermion, Majorana masses of right-handed (sterile) neutrinos in this model should be excluded in the interval $10~\text{GeV} \lesssim m_\psi \lesssim 60~\text{GeV}$.  

Of course, couplings of the form \eqref{sterile} will also contribute to the naturalness issue for the Higgs-boson mass (see \cite{Dine:2015xga} for a review of the naturalness issues of the standard Model). 

\section{Discussion}
\label{sec:discuss}

We have shown that a scale-invariant theory with Lagrangian of the form \eqref{L} exhibits dynamical breaking of scale invariance, hence, also of electroweak symmetry.  This is best seen in terms of variables of \eqref{L1}, in which the Higgs-field potential has an absolute minimum corresponding to $\Phi^\dagger \Phi = \chi^2$ while the auxiliary field $\chi$ can take any non-zero value due to scale invariance.  In the original frame \eqref{L}, it is the vacuum Higgs-field invariant $\Phi^\dagger \Phi$ and scalar curvature that can take any value, related by \eqref{pr}, due to scale invariance of the theory.  The term $\lambda_\Phi \left( \Phi^\dagger \Phi \right)^2$ in \eqref{L} or the corresponding term $\lambda_\chi \chi^4$ in \eqref{L1} generate a small cosmological constant fixing the background spacetime metric.  The naturalness issue related to the smallness of the gravitational and cosmological constants is translated as the issue of largeness of the constant $\xi_* \approx 10^{32}$ and smallness of $\lambda_\Phi \approx 3 \times 10^{-56}$ in \eqref{L}.

Note that the theory does not have a continuous limit of $\lambda_* \to \infty$, i.e., to the case where the $R^2$ term is absent from \eqref{L}.  Indeed, in this case, the gravitational scalaron degree of freedom disappears, and the theory becomes the one considered in \cite{Zee:1978wi, vanderBij:1993hx, vanderBij:1994bv}. Conformal rescaling of the metric is then performed by using the Higgs-field invariant $\Phi^\dagger \Phi$, and the Higgs boson in this case becomes massless and decouples from the Standard Model. In the presence of $R^2$ term, as $\lambda_* \to \infty$ (implying $\lambda \to \infty$), according to \eqref{Hd}, the Higgs boson becomes infinitely heavy, leading to a massive Yang--Mills theory of vector bosons.

Along with a fourth-order gravity term quadratic in the scalar curvature, one can add to \eqref{L} another fourth-order term proportional to $C_{\alpha\beta\mu\nu} C^{\alpha\beta\mu\nu}$, where $C_{\alpha\beta\mu\nu}$ is the conformal Weyl tensor. Since this term in the action is Weyl-invariant, it will remain unmodified in the final result \eqref{L2}, leading to a version of Stelle gravity \cite{Stelle:1976gc, Stelle:1977ry}. Such a theory is plagued with ghosts, with a possible resolution of this problem consisting in allowing for curvature invariants of unlimited differential order in the action \cite{Kuntz:2019qcf}. Hopefully, this can be done in a scale-invariant manner by using the Higgs field, without affecting qualitatively the lower-order behaviour considered here.

We treated Lagrangian \eqref{L} as an \emph{effective\/} Lagrangian arising in a scale-invariant quantum field theory involving gravity. In order to maintain scale-invariance on the quantum level, one should assume some scale-invariant prescription for regularisation, e.g., employing the $\chi$-dependent regularisation parameter \cite{Englert:1976ep, Shaposhnikov:2008xi} when dealing with \eqref{L1}. In the present case, $\chi$-dependence from the viewpoint of the original Lagrangian \eqref{L} means dependence on the combination 
\begin{equation}
\chi_*^2 = \Phi^\dagger \Phi + \frac{\xi_*}{2 \lambda_*} R \, ,
\end{equation}
obtained by variation of \eqref{L0} with respect to $\chi_*^2$.  To preserve the structure of \eqref{L0}, one will need to ensure that no kinetic term for the field $\chi$ arises in regularising \eqref{L1}. This regularisation prescription means the usual field-independent regularisation of \eqref{L2} preserving the relations between the terms which involve the constant $M$. 

In passing from \eqref{L1} to \eqref{L2}, a Weyl transformation of fields was made. On the quantum level, this might produce anomalous terms, whose origin can be traced to the transformation of the quantum measure in the path integral. The quantum theory is assumed to preserve the global scale-invariance; hence, the anomalous terms in the Einstein frame should respect the shift symmetry $\phi \to \phi + \text{const}$ of the action up to boundary terms. Presumably, then, they will contain terms such as $\left( \phi / M \right) L_\text{\tiny GB}$, where $L_\text{\tiny GB}$ is the Gauss--Bonnet term (Euler density). This depends on the specification of the path-integral measure and regularisation prescription preserving the global scale-invariance, which requires further elaboration.

The model under consideration does not describe successful inflation based on the modified Higgs field. This is because the coupling constant $\lambda \approx 0.13$ in \eqref{Hd} is way too large to generate the observable initial power spectrum. In extending the model appropriately, one can preserve scale-invariance, but one also needs to ensure that no overproduction of the dilaton radiation occurs in the reheating process. This does not seem to be problematic given that the dilaton interacts directly\,---\,and very weakly\,---\,only with the Higgs field by means of its coupling to the scalar curvature in the original frame \eqref{L}.

\acknowledgments

I am grateful to Massimiliano Rinaldi for valuable communication. The author acknowledges support from the Simons Foundation.  This work is supported by the National Academy of Sciences of Ukraine under project 0121U109612 and by the Taras Shevchenko National University of Kyiv under project 22BF023-01.


\begin{thebibliography}{9}

\bibitem{Fujii:1974bq}
Y.~Fujii, 
\emph{Scalar-tensor theory of gravitation and spontaneous breakdown of scale invariance}, 
\href{https://doi.org/10.1103/PhysRevD.9.874}{\emph{Phys. Rev. D} \textbf{9} (1974) 874--876}
[\href{https://inspirehep.net/literature/93041}{inSPIRE}].

\bibitem{Fujii:1982ms}
Y.~Fujii, 
\emph{Origin of the gravitational constant and particle masses in a scale-invariant scalar-tensor theory}, 
\href{https://doi.org/10.1103/PhysRevD.26.2580}{\emph{Phys. Rev. D} \textbf{26} (1982) 2580--2588} 
[\href{https://inspirehep.net/literature/178794}{inSPIRE}].

\bibitem{Englert:1975wj}
F.~Englert, E.~Gunzig, C.~Truffin and P.~Windey,
\emph{Conformal invariant general relativity with dynamical symmetry breakdown}, 
\href{https://doi.org/10.1016/0370-2693(75)90247-6}{\emph{Phys. Lett. B} \textbf{57} (1975) 73--77}
[\href{https://inspirehep.net/literature/2115}{inSPIRE}].

\bibitem{Englert:1976ep}
F.~Englert, C.~Truffin and R.~Gastmans, 
\emph{Conformal invariance in quantum gravity}, 
\href{https://doi.org/10.1016/0550-3213(76)90406-5}{\emph{Nucl. Phys. B} \textbf{117} (1976) 407--432}
[\href{https://inspirehep.net/literature/3485}{inSPIRE}].

\bibitem{Wetterich:1987fk}
C.~Wetterich, 
\emph{Cosmologies with variable Newton's ``constant''},
\href{https://doi.org/10.1016/0550-3213(88)90192-7}{\emph{Nucl. Phys. B} \textbf{302} (1988) 645--667}
[\href{https://inspirehep.net/literature/22489}{inSPIRE}].

\bibitem{Wetterich:1987fm}
C.~Wetterich, 
\emph{Cosmology and the fate of dilatation symmetry},
\href{https://doi.org/10.1016/0550-3213(88)90193-9}{\emph{Nucl. Phys. B} \textbf{302} (1988) 668--696}
[\href{https://arxiv.org/abs/1711.03844}{arXiv:1711.03844}] [\href{https://inspirehep.net/literature/22490}{inSPIRE}].

\bibitem{Shaposhnikov:2008xb}
M.~Shaposhnikov and D.~Zenh\"{a}usern, 
\emph{Scale invariance, unimodular gravity and dark energy},
\href{https://doi.org/10.1016/j.physletb.2008.11.054}{\emph{Phys. Lett. B} \textbf{671} (2009) 187--192}
[\href{https://arxiv.org/abs/0809.3395}{arXiv:0809.3395}]
[\href{https://inspirehep.net/literature/797084}{inSPIRE}].

\bibitem{Shaposhnikov:2008xi}
M.~Shaposhnikov and D.~Zenh\"{a}usern,
\emph{Quantum scale invariance, cosmological constant and hierarchy problem},
\href{https://doi.org/10.1016/j.physletb.2008.11.041}{\emph{Phys. Lett. B} \textbf{671} (2009) 162--166}
[\href{https://arxiv.org/abs/0809.3406}{arXiv:0809.3406}]
[\href{https://inspirehep.net/literature/797091}{inSPIRE}].

\bibitem{Gretsch:2013ooa}
F.~Gretsch and A.~Monin,
\emph{Perturbative conformal symmetry and dilaton},
\href{https://doi.org/10.1103/PhysRevD.92.045036}{\emph{Phys. Rev. D} \textbf{92}, no.~4 (2015) 045036}
[\href{https://arxiv.org/abs/1308.3863}{arXiv:1308.3863}]
[\href{https://inspirehep.net/literature/1249626}{inSPIRE}].

\bibitem{Shaposhnikov:2022dou}
M.~Shaposhnikov and A.~Tokareva,
\emph{Anomaly-free scale symmetry and gravity},
\href{https://doi.org/10.1016/j.physletb.2023.137898}{\emph{Phys. Lett. B} \textbf{840} (2023) 137898}
[\href{https://arxiv.org/abs/2201.09232}{arXiv:2201.09232}]
[\href{https://inspirehep.net/literature/2016467}{inSPIRE}].

\bibitem{Shaposhnikov:2022zhj}
M.~Shaposhnikov and A.~Tokareva,
\emph{Exact quantum conformal symmetry, its spontaneous breakdown, and gravitational Weyl anomaly},
\href{https://doi.org/10.1103/PhysRevD.107.065015}{\emph{Phys. Rev. D} \textbf{107} (2023) 065015}
[\href{https://arxiv.org/abs/2212.09770}{arXiv:2212.09770}]
[\href{https://inspirehep.net/literature/2616328}{inSPIRE}].

\bibitem{Wetterich:2019qzx}
C.~Wetterich,
\emph{Quantum scale symmetry},
\href{https://arxiv.org/abs/1901.04741}{arXiv:1901.04741}
[\href{https://inspirehep.net/literature/1713827}{inSPIRE}]. 

\bibitem{Brans:1961sx}
C.~Brans and R.~H.~Dicke,
\emph{Mach's principle and a relativistic theory of gravitation},
\href{https://doi.org/10.1103/PhysRev.124.925}{\emph{Phys. Rev.} \textbf{124} (1961) 925--935}
[\href{https://inspirehep.net/literature/2450}{inSPIRE}].

\bibitem{Dicke:1961gz}
R.~H.~Dicke,
\emph{Mach's principle and invariance under transformation of units},
\href{https://doi.org/10.1103/PhysRev.125.2163}{\emph{Phys. Rev.} \textbf{125} (1962) 2163--2167}
[\href{https://inspirehep.net/literature/2035}{inSPIRE}].

\bibitem{Zee:1978wi}
A.~Zee,
\emph{Broken-Symmetric Theory of Gravity},
\href{https://doi.org/10.1103/PhysRevLett.42.417}{\emph{Phys. Rev. Lett.} \textbf{42} (1979) 417--421}
[\href{https://inspirehep.net/literature/133157}{inSPIRE}].

\bibitem{vanderBij:1993hx}
J.~J.~van der Bij,
\emph{Can gravity make the Higgs particle decouple?},
{\emph{Acta Phys. Polon. B} \textbf{25} (1994) 827--832}
[\href{https://inspirehep.net/literature/360199}{inSPIRE}].

\bibitem{vanderBij:1994bv}
J.~J.~van der Bij,
\emph{Can gravity play a role at the electroweak scale?},
\emph{Int. J. Phys.} \textbf{1} (1995) 63
[\href{https://arxiv.org/abs/hep-ph/9507389}{hep-ph/9507389}]
[\href{https://inspirehep.net/literature/381722}{inSPIRE}].

\bibitem{Gorbunov:2013dqa}
D.~Gorbunov and A.~Tokareva,
\emph{Scale-invariance as the origin of dark radiation?},
\href{https://doi.org/10.1016/j.physletb.2014.10.036}{\emph{Phys. Lett. B} \textbf{739} (2014) 50--55}
[\href{https://arxiv.org/abs/1307.5298}{arXiv:1307.5298}]
[\href{https://inspirehep.net/literature/1243583}{inSPIRE}].

\bibitem{Rinaldi:2015uvu}
M.~Rinaldi and L.~Vanzo,
\emph{Inflation and reheating in theories with spontaneous scale invariance symmetry breaking},
\href{https://doi.org/10.1103/PhysRevD.94.024009}{\emph{Phys. Rev. D} \textbf{94}, no.2 (2016) 024009}
[\href{https://arxiv.org/abs/1512.07186}{arXiv:1512.07186}]
[\href{https://inspirehep.net/literature/1411046}{inSPIRE}].

\bibitem{Tambalo:2016eqr}
G.~Tambalo and M.~Rinaldi,
\emph{Inflation and reheating in scale-invariant scalar-tensor gravity},
\href{https://doi.org/10.1103/PhysRevD.94.024009}{\emph{Gen. Rel. Grav.} \textbf{49}, no.~4 (2017) 52}
[\href{https://arxiv.org/abs/1610.06478}{arXiv:1610.06478}]
[\href{https://inspirehep.net/literature/1492955}{inSPIRE}].

\bibitem{Ghoshal:2022qxk}
A.~Ghoshal, D.~Mukherjee and M.~Rinaldi,
\emph{Inflation and primordial gravitational waves in scale-invariant quadratic gravity with Higgs},
\href{https://doi.org/10.1007/JHEP05(2023)023}{\emph{JHEP} \textbf{05} (2023) 023}
[\href{https://arxiv.org/abs/2205.06475}{arXiv:2205.06475}]
[\href{https://inspirehep.net/literature/2081863}{inSPIRE}].

\bibitem{Ghilencea:2018dqd}
D.~M.~Ghilencea,
\emph{Spontaneous breaking of Weyl quadratic gravity to Einstein action and Higgs potential},
\href{https://doi.org/10.1007/JHEP03(2019)049}{\emph{JHEP} \textbf{03} (2019) 049}
[\href{https://arxiv.org/abs/1812.08613}{arXiv:1812.08613}]
[\href{https://inspirehep.net/literature/1710395}{inSPIRE}].

\bibitem{Ghilencea:2021lpa}
D.~M.~Ghilencea,
\emph{Standard Model in Weyl conformal geometry},
\href{https://doi.org/10.1140/epjc/s10052-021-09887-y}{\emph{Eur. Phys. J. C} \textbf{82}, issue~1 (2022) 23}
[\href{https://arxiv.org/abs/2104.15118}{arXiv:2104.15118}]
[\href{https://inspirehep.net/literature/1861597}{inSPIRE}].

\bibitem{Shtanov:1994sh}
Y.~V.~Shtanov and S.~A.~Yushchenko,
\emph{Conformally invariant cosmology based on Riemann--Cartan space-time},
\href{https://doi.org/10.1088/0264-9381/11/10/007}{\emph{Class. Quant. Grav.} \textbf{11} (1994) 2455--2482}
[\href{https://arxiv.org/abs/gr-qc/9402033}{gr-qc/9402033}]
[\href{https://inspirehep.net/literature/37352}{inSPIRE}].

\bibitem{Penrose:1987}
R.~Penrose and W.~Rindler, \emph{Spinors and Space-Time. Volume~1: Two-spinor calculus and relativistic fiels}, Cambridge University Press, Cambridge U.K. (1987), p.~458.

\bibitem{LHCHiggs:2016ypw}
\textsc{LHC Higgs Cross Section Working Group} collaboration,
\emph{Handbook of LHC Higgs Cross Sections: 4. Deciphering the Nature of the Higgs Sector}, 
\href{https://arxiv.org/abs/1610.07922}{arXiv:1610.07922}
[\href{https://doi.org/10.23731/CYRM-2017-002}{DOI:10.23731/CYRM-2017-002}]
[\href{https://inspirehep.net/literature/1494411}{inSPIRE}].

\bibitem{ATLAS:2023dnm}
\textsc{ATLAS} collaboration,
\emph{Evidence of off-shell Higgs boson production from $ZZ$ leptonic decay channels and constraints on its total width with the ATLAS detector}, 
\href{https://doi.org/10.1016/j.physletb.2023.138223}{\emph{Phys. Lett. B} \textbf{846} (2023) 138223}
[\href{https://arxiv.org/abs/2304.01532}{arXiv:2304.01532}]
[\href{https://inspirehep.net/literature/2648835}{inSPIRE}].

\bibitem{Stelle:1976gc}
K.~S.~Stelle,
\emph{Renormalization of higher-derivative quantum gravity},
\href{https://doi.org/10.1103/PhysRevD.16.953}{\emph{Phys. Rev. D} \textbf{16} (1977) 953--969}
[\href{https://inspirehep.net/literature/110537}{inSPIRE}].

\bibitem{Stelle:1977ry}
K.~S.~Stelle,
\emph{Classical gravity with higher derivatives},
\href{https://doi.org/10.1007/BF00760427}{\emph{Gen. Rel. Grav.} \textbf{9} (1978) 353--371}
[\href{https://inspirehep.net/literature/119488}{inSPIRE}].

\bibitem{Kuntz:2019qcf}
I.~Kuntz,
\emph{Exorcising ghosts in quantum gravity},
\href{https://doi.org/10.1140/epjp/s13360-020-00875-x}{\emph{Eur. Phys. J. Plus} \textbf{135}, no.~10 (2020) 859}
[\href{https://arxiv.org/abs/1909.11072}{arXiv:1909.11072}]
[\href{https://inspirehep.net/literature/1755691}{inSPIRE}].

\bibitem{Dine:2015xga}
M.~Dine,
\emph{Naturalness Under Stress},
\href{https://doi.org/10.1146/annurev-nucl-102014-022053}{\emph{Ann. Rev. Nucl. Part. Sci.} \textbf{65} (2015) 43--62}
[\href{https://arxiv.org/abs/1501.01035}{arXiv:1501.01035}]
[\href{https://inspirehep.net/literature/1336631}{inSPIRE}].

\end{thebibliography}
\end{document}